\documentclass[aps,pra,floatfix,twocolumn,showpacs,10pt]{revtex4-1}
\usepackage{amssymb, amsmath,color,mciteplus,graphicx,subfigure}
\usepackage{calc,epsfig,epstopdf,color,mciteplus,bm,mathrsfs}
\usepackage{times}
\usepackage{float}
\usepackage{lipsum}

\begin{document}

\title{Nonadiabatic geometric quantum computation with parametrically tunable coupling}

\author{Tao Chen}

\author{Zheng-Yuan Xue}\email{zyxue83@163.com}
\affiliation{Guangdong Provincial Key Laboratory of Quantum Engineering and Quantum Materials, and School of Physics\\ and Telecommunication Engineering, South China Normal University, Guangzhou 510006, China}

\date{\today}
\begin{abstract}
The nonadiabatic geometric quantum computation is promising as it is robust against certain types of local noises. However, its experimental implementation is challenging due to the need of complex control on  multi-level and/or multiple quantum systems. Here, we propose to implement it on a two-dimensional square superconducting qubit lattice. In the construction of our geometric quantum gates, we only use the simplest and experimentally accessible control over the qubit states of the involved quantum systems, without introducing any auxiliary state. Specifically, our scheme is achieved by parametrically tunable all-resonant interaction, which leads to high-fidelity  quantum gates. Meanwhile, this simple implementation can be conveniently generalized to a composite scenario, which can further suppress the systematic error during the gate operations. In addition, universal nonadiabatic geometric quantum gates in decoherence-free subspace can also be realized based on the tunable coupling between only two transmon qubits, without consulting to multiple qubits and only using two physical qubits to encode a logical qubit. Therefore, our proposal provides a promising way of high-fidelity geometric manipulation for robust solid-state quantum computation.
\end{abstract}

\maketitle

\section{Introduction}

The superiority of quantum computation is generally believed to be mainly reflected by the manipulation of the superposition of quantum states, which can solve some certain problems that are difficult to process by classical computation \cite{Shor}. However, the manipulation of quantum states is susceptible to the noises induced by their surrounding environment, which makes the coherence time of the target quantum systems limited. Meanwhile, scalable quantum computation needs high-precision quantum manipulation. Therefore, how to obtain high-fidelity quantum gates is the key towards large-scale fault-tolerant quantum computation.

Superconducting quantum circuits \cite{sq1,sq2,sq3}, which have demonstrated long coherent time, high-fidelity quantum gates, and reliable readout techniques, have been shown to be promising in the physical implementation of quantum computation. To realize high-fidelity quantum gates, coupling between qubits should be able to be selectively tuned, which is difficult for solid-state qubits as the interactions are of the always-on nature. Generally, there are mainly three different ways towards tunable coupling between superconducting qubits. Firstly, for qubits with fixed frequencies, transverse microwave drives can activate multiqubit interactions \cite{TMDrive1,TMDrive2}. However, in this case, stringent restrictions on the frequencies and anharmonicities of the qubits are imposed. Secondly, for qubits with tunable frequencies, two-qubit gates can be obtained when they are tuned into resonance \cite{DiCarlo09,Kelly15}. But, tuning a qubit frequency will also result in shorter coherent time. In addition, for both cases, due to the limited qubit anharmonicity, unwanted cross talk, with energy levels beyond the qubit levels, limits the scale up of the qubit lattice. Thirdly, tunable coupling between two qubits with different frequencies can be obtained by modulating one of the qubits or the coupling between them at a frequency equal to their detuning \cite{TC1,TC2,TC3,TC0,TC4,TC5,TC6,TC7,TC8}, i.e., the parametrically tunable coupling, where the modulating field can provide additional control over the coupling strength. In this case, a modulating field can be added when a qubit works on its optimal frequency point \cite{TC6}, and thus does not cause the shortening of qubit coherent times. Therefore, combined with its selective implementation, this parametrically tunable coupling can mitigate the frequency crowding problem when scaling up the qubit lattice.

To implement high-fidelity quantum gates, geometric quantum computation (GQC) schemes \cite{AGQC1,Duan,AGQC2,AGQC3,AGQC4,AGQC5,AGQC6} have been proposed, based on both Abelian \cite{Abelian} and non-Abelian geometric phases \cite{non-Abelian}, where the quantum gates only depend on the global properties of the evolution paths so that they are robust against certain local noises. However, the GQC is originally proposed by using the adiabatic cycle evolution, where quantum systems will be exposed to the external environment for a long time. To remove this adiabatic limitation, nonadiabatic GQC schemes have been proposed to achieve fast and high-fidelity quantum gates based on Abelian \cite{wxb,ZSL1,ZSL2,NGQC} and non-Abelian geometric phases  \cite{TongDM,ZhangJ,Singleshot,Singleloop, Composite,Dressedstate,3level}, which can still have the merit of robustness against certain local noises \cite{stochastic1,robust1,stochastic2,robust2,robust3}. Recently, geometric phases or elementary geometric quantum gates have been experimentally demonstrated on various systems \cite{exp1,exp2,exp3,exp6,exp4,exp5,Abdumalikov35,ibmexp36,xuy37,sust38,Feng39,li40,Zu41, Arroyo-Camejo42,nv2016,nv43,nv44}. However, universal GQC is still experimentally difficult due to the need for complex control over qubit-qubit and/or qubit-bus interactions, where auxiliary energy levels beyond the qubit states and/or auxiliary coupling elements may be needed.

In addition, when a qubit interacts with its environment, qubit dephasing noise will be produced, which is one of the major sources of qubit decoherence. However, for many qubit system, this type of noise may behave collectively \cite{DFS1,DFS2,DFS3}. To suppress such noise, nonadiabatic GQC in decoherence-free subspace (DFS) based on Abelian \cite{fengxl,DFS5,DFS8} and non-Abelian geometric phases \cite{DFS4,DFS6,DFS7,DFS9} has been proposed to further improve the precision of the quantum gate control. Again, these realizations of the GQC in DFS are experimentally challenging as they usually need delicate interactions among multiple quantum systems.

Here, we propose to implement nonadiabatic GQC (as well as in DFS) on a two-dimensional (2D) square superconducting qubit lattice. In the construction of our universal geometric quantum gates, we merely use simple and experimentally accessible microwave control over two states of the involved capacitively coupled superconducting transmon qubits, where the leakage of qubit states can be effectively suppressed. Remarkably, our scheme can be achieved by parametrically tunable coupling between only two qubits, thus leading to high-fidelity geometric quantum gates in an all-resonant way. Meanwhile, our implementation can be directly generalized to a composite scenario, which can further suppress the systematic error during the gate operations. In addition, we only use two physical qubits to construct the DFS, which requires the minimal qubit resource. Therefore, our scheme provides a promising way to achieve high-fidelity GQC on a large-scale qubit lattice.

\begin{figure}[tbp]
  \centering
  \includegraphics[width=0.98\linewidth]{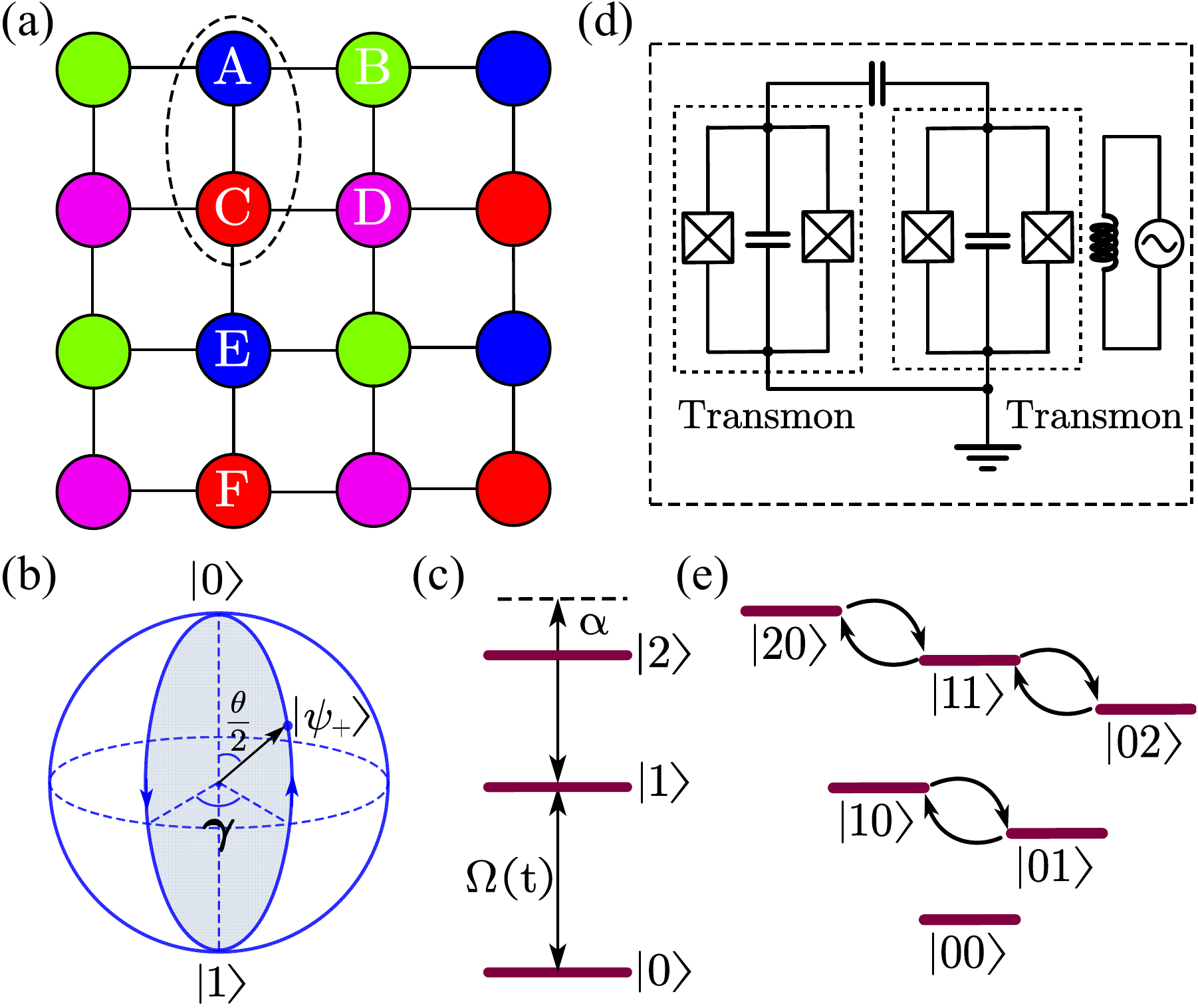}
  \caption{Illustration of the proposed setup of our scheme. (a) The considered 2D square qubit lattice consists of capacitively coupled superconducting transmon qubits, with different colors denoting qubits with different frequencies. (b) The orange-slice-shaped evolution path, depicted in a Bloch sphere, chose to induce the wanted arbitrary geometric phase $\gamma$. (c) Energy levels of a superconducting transmon, the two lowest levels of which are used as our qubit states. Due to the weak anharmonicity ($\alpha$) nature of the transmon qubit, external microwave field driving can also induce the dispersive transitions between the higher excited states, which causes leakage of the encoded quantum information. The two  transmon qubits marked with a ellipse in (a) represent two capacitively coupled qubits, and the details are shown in (d), where one of them is biased by an ac magnetic flux, which periodically modulates its transition frequency. Meanwhile, these two coupled qubits can be treat as a unit to encode a logical qubit for a decoherence-free subspace encoding. (e) Energy level diagram for two capacitively coupled transmon qubits, where single- and two-excitation subspaces can be used to implement the geometric iSWAP and control-phase gates, respectively.}\label{Fig1}
\end{figure}

\section{Universal single-qubit geometric gates}

\subsection{Geometric gates}

Here, we propose to implement a nonadiabatic GQC scheme on the 2D square qubit lattice consisting of capacitively coupled superconducting transmon qubits \cite{Transmon,jqyou}, as shown in Fig. \ref{Fig1}(a). This 2D configuration verifies that our proposal can support large-scale universal quantum computation. We first proceed to the implementation of the universal nonadiabatic single-qubit geometric gates, by coupling the two lowest levels of a transmon qubit with a microwave field resonantly. In order to clearly explain the geometric nature of the implemented gate, we firstly consider the ideal situation without leakage. Assuming $\hbar=1$ hereafter, the reduced Hamiltonian in the computational basis $|0\rangle$ and $|1\rangle$ can be written as
\begin{eqnarray}
\label{Eq1}
H_1(t)=
\frac {1} {2}\Omega(t) \left(
\begin{array}{cccc}
 0 & e^{-i\phi} \\
 e^{i\phi} & 0
\end{array}
\right),
\end{eqnarray}
where $\Omega(t)$ and $\phi$ are the driving strength and phase of the microwave field, respectively.

To get a geometric evolution, we divide the entire evolution time $\mathrm{T}$ into three parts, at the intermediate time $\mathrm{T}_1$, $\mathrm{T}_2$ with pulse area and relative phase $\phi$ satisfying
\begin{eqnarray}
\label{Eq2}
\int^{\mathrm{T}_1}_0 \Omega(t)dt&=&\theta,               \quad  \phi-\frac {\pi} {2}, \quad \quad \quad t\in[0,\mathrm{T}_1],              \notag\\
\int^{\mathrm{T}_2}_{\mathrm{T}_1} \Omega(t)dt&=&\pi,     \quad  \phi+\gamma+\frac {\pi} {2},      \quad t\in[\mathrm{T}_1,\mathrm{T}_2],   \notag\\
\int^{\mathrm{T}}_{\mathrm{T}_2} \Omega(t)dt&=&\pi-\theta,\quad  \phi-\frac {\pi} {2},             \quad t\in[\mathrm{T}_2,\mathrm{T}].
\end{eqnarray}
At the final time $\mathrm{T}$, the evolution operator can be expressed as
\begin{eqnarray}
\label{Eq3}
U_1(\mathrm{T})&=&U_1(\mathrm{T},\mathrm{T}_2) U_1(\mathrm{T}_2,\mathrm{T}_1) U_1(\mathrm{T}_1,0) \notag\\
&=&\cos{\gamma}+i\sin{\gamma}\left(
\begin{array}{cccc}
 \cos{\theta} & \sin{\theta}e^{-i\phi} \\
 \sin{\theta}e^{i\phi} & -\cos{\theta}
\end{array}
\right)\notag\\
&=&e^{i\gamma \vec{n}\cdot\vec{\sigma} }
\end{eqnarray}
where $\vec{n}=(\sin{\theta}\cos{\phi},\sin{\theta}\sin{\phi},\cos{\theta})$, the parameters $\theta,\phi$ and $\gamma$ can be tuned by external microwave field control, and $\vec{\sigma}=(\sigma_x,\sigma_y,\sigma_z)$ are the Pauli operators for the computational subspace $\{|0\rangle, |1\rangle\}$.

Then, we demonstrate that $U_1(\mathrm{T})$ is a geometric gate \cite{NGQC}. We take the two-dimensional orthogonal eigenstates
\begin{eqnarray}
\label{Eq4}
\left\{
\begin{array}{ll}
|\psi_+\rangle=\cos{\frac {\theta} {2}}|0\rangle+\sin{\frac {\theta} {2}}e^{i\phi}|1\rangle, \\
\ \ \\
|\psi_-\rangle=\sin{\frac {\theta} {2}}e^{-i\phi}|0\rangle-\cos{\frac {\theta} {2}}|1\rangle, \\
\end{array}
\right.
\end{eqnarray}
of $\vec{n}\cdot\vec{\sigma}$, a pair of dressed states, as our evolution states inducing the geometric gate. In this dressed-state representation, the evolution operator can be rewritten as
\begin{eqnarray}
\label{Eq5}
U_1(\mathrm{T})=e^{i\gamma}|\psi_+\rangle\langle \psi_+|+e^{-i\gamma}|\psi_-\rangle\langle \psi_-|.
\end{eqnarray}
Clearly, both of the orthogonal states $|\psi_+\rangle$ and $|\psi_-\rangle$ satisfy the cyclic evolution condition, i.e.,
\begin{eqnarray}
\label{Eq6}
|\psi_+(\mathrm{T})\rangle&=&U_1(\mathrm{T})|\psi_+\rangle=e^{i\gamma}|\psi_+\rangle, \notag\\
|\psi_-(\mathrm{T})\rangle&=&U_1(\mathrm{T})|\psi_-\rangle=e^{-i\gamma}|\psi_-\rangle,
\end{eqnarray}
and the parallel-transport condition, i.e.,
\begin{eqnarray}
\label{Eq7}
\langle \psi_\pm|U_1^\dag(t)H_1(t)U_1(t)|\psi_\pm\rangle=0.
\end{eqnarray}
Therefore, after undergoing a single orange-slice-shaped cyclic evolution path, as shown in Fig. \ref{Fig1}(b), at the evolution time $\mathrm{T}$,  $|\psi_\pm\rangle$ can obtain pure geometric phases $\pm\gamma$ without any dynamic phase. Therefore, universal nonadiabatic single-qubit geometric gates in Eq. (\ref{Eq3}) can be achieved.

\subsection{Gate performance}

\begin{figure}[tbp]
  \centering
  \includegraphics[width=0.9\linewidth]{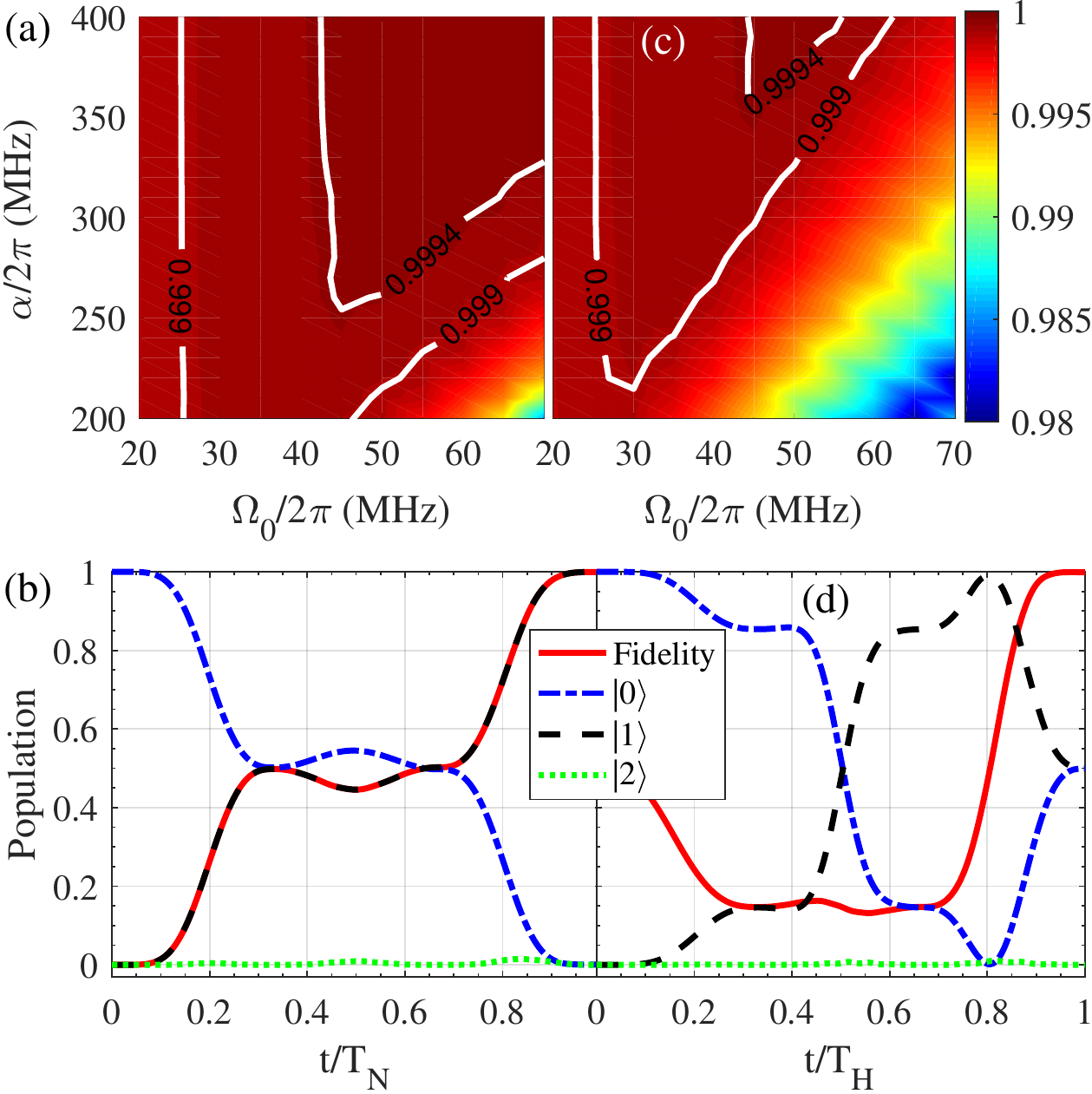}
  \caption{Gate fidelities as functions of the driving amplitude $\Omega_0$ and anharmonicity $\alpha$ of the transmon qubit, the results of the NOT and Hadamard gates are shown in (a) and (c), respectively. When the initial state being $|0\rangle$, the qubit-state population and the state-fidelity dynamics of the NOT and Hadamard gate operations  are shown in (b) and (d), respectively.}
  \label{Fig2}
\end{figure}

Consider the intrinsic weak anharmonicity $\alpha$ of the transmon qubit, the microwave driving will also couple the high energy-level in a dispersive way, as shown in Fig. \ref{Fig1}(c). Here, we take the simple pulse shape $\Omega(t)=\Omega_0 \sin^2\left(\pi t/\mathrm{T}\right)$ as an example and use the demonstrated DRAG correction to suppress this leakage (see Appendix A for details), so that one can obtain high-fidelity geometric  manipulation on the transmon qubit states. To analyze the performance of the single-qubit gates, we choose the NOT and Hadamard gates as two typical examples which correspond to the same $\phi=0$ and $\gamma=\pi/2$, with different $\theta$, $\theta_N=\pi/2$ and $\theta_H=\pi/4$ for the NOT and Hadamard gates, respectively. We numerically simulate the gate performance by using the master equation (see Appendix A for details). In our simulation, we choose the parameters from the state-of-art experiments in Refs. \cite{DRAGexperiment1,DRAGexperiment2}, that is $\kappa^1_-=\kappa^1_z=\frac {1} {2} \kappa^2_-=\frac {1} {2} \kappa^2_z=\kappa=2\pi\times 4$ KHz. For a general initial state $|\psi_1 \rangle=\cos\theta_1|0\rangle+\sin\theta_1|1\rangle$, the NOT and Hadamard gates should result in an ideal final state $|\psi_{f_N}\rangle=\cos\theta_1|1\rangle+\sin\theta_1|0\rangle$ and
$|\psi_{f_H}\rangle=\frac {1} {\sqrt2} [(\cos\theta_1+\sin\theta_1)|0\rangle+(\cos\theta_1-\sin\theta_1)|1\rangle]$. To fully evaluate the performance of the implemented gates, we define the gate fidelity as $F_{N/H}^G=\frac {1} {2\pi}\int_0^{2\pi} \langle  \psi_{f_{N/H}}|\rho_1|\psi_{f_{N/H}}\rangle d\theta_1$ \cite{Gatefidelity} with the integration numerically performed for 1001 input states with $\theta_1$ being uniformly distributed over $[0, 2\pi]$. In Fig. \ref{Fig2}(a) and (c), we plot the gate fidelities as functions of the driving amplitude $\Omega_0$ and anharmonicity $\alpha$ of the transmon qubit, where we find that the gate fidelities of the NOT and Hadamard gates can, respectively, reach $99.95\%$ and $99.94\%$ for a certain range of parameters (within the current experimental reach).

Furthermore, our geometric implementation of a single-qubit gate has the same level of gate fidelities as that of the dynamical method, the best performance of which is reported in Refs. \cite{DRAGexperiment1,DRAGexperiment2}. In the following, we use the experimental parameters reported there as a basis for comparison, i.e., we set the anharmonicity as $\alpha=2\pi\times220$ MHz. Suppose the qubit is initially in the state $|\psi_1\rangle=|0\rangle$, the NOT and Hadamard gates should result in the ideal final states $|\psi_{f_N}\rangle=|1\rangle$ and $|\psi_{f_H}\rangle=(|0\rangle+|1\rangle)/\sqrt{2}$, respectively. We evaluate these gates by state populations and the state fidelities defined by $F_{N/H}=\langle\psi_{f_{N/H}}|\rho_1|\psi_{f_{N/H}}\rangle$. In this way, we obtain a very high state fidelity $F_N=99.93\%$ with $\Omega_0=2\pi\times40$ MHz and $F_H=99.89\%$ with $\Omega_0=2\pi\times30$ MHz for the NOT and Hadamard geometric gates, as shown in Fig. \ref{Fig2}(b) and (d), where the corresponding gate fidelities are $99.93\%$ and $99.91\%$. We have numerically verified that the infidelity that results from the error caused by coupling to adjacent transmon qubits is less than 0.01\%, when the qubit-frequency difference of adjacent transmon qubits is larger than $2\pi\times1$ GHz. This is also valid for the two-qubit interaction case, so we ignore this type of error hereafter.

\begin{figure}[tbp]
  \centering
  \includegraphics[width=0.95\linewidth]{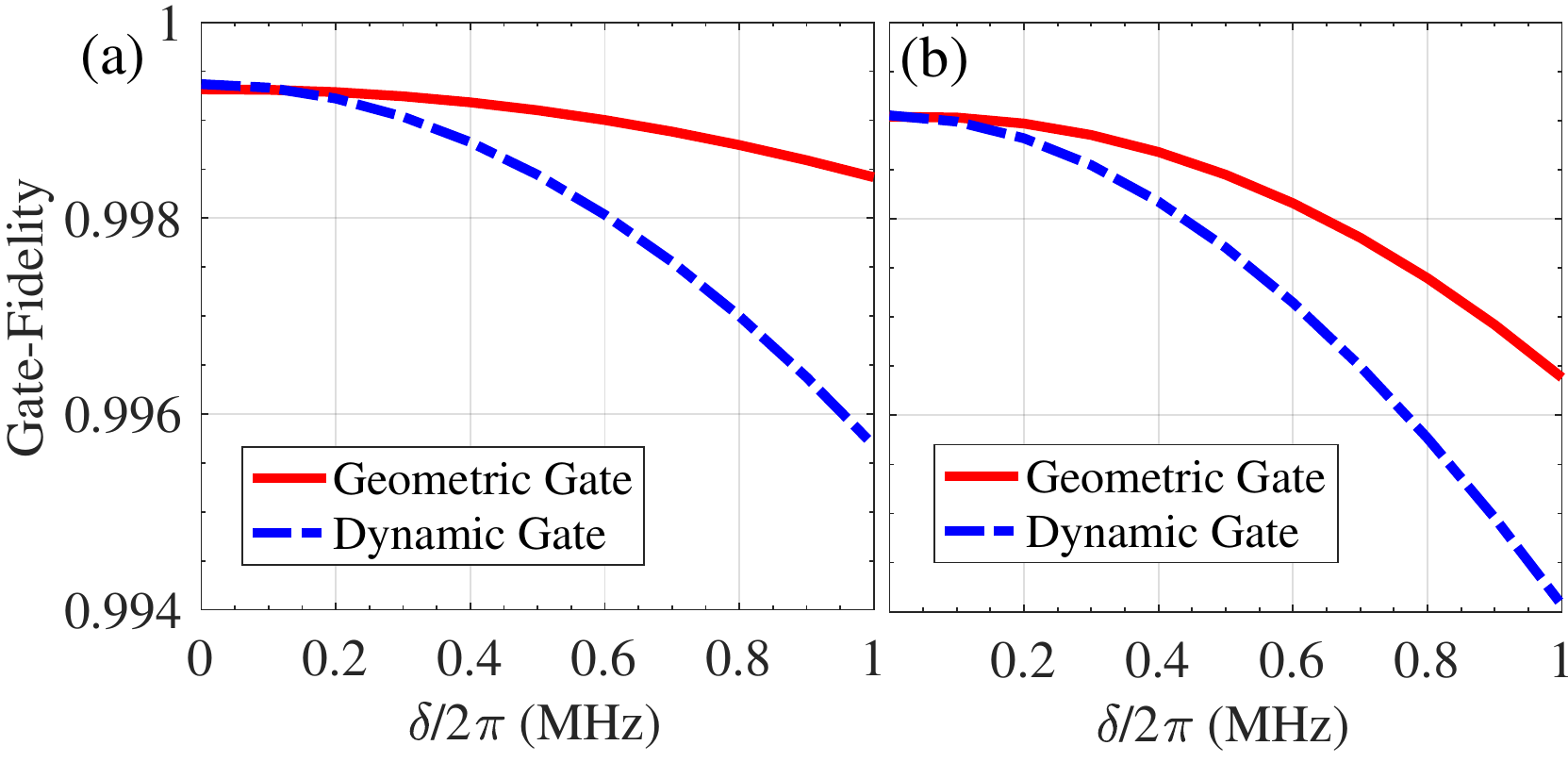}
  \caption{Gate fidelities for geometric and dynamic
  (a) NOT and (b) Hadamard gates as a function of the qubit-frequency drift $\delta$.}
  \label{Fig3}
\end{figure}

To show the noise-resilient feature of our geometric gates, we take the error caused by the qubit frequency drift as an example to compare the performance of our geometric gates with that of the corresponding dynamic gates. In our simulation, we set the operation time of both the dynamic and geometric gates to be the same. Taking the third level into account, the whole Hamiltonian consists of Eq. (\ref{EqA1}) plus an additional error term $(\delta |1\rangle\langle 1|+2\delta |2\rangle\langle 2|)$ due to the qubit-frequency drift with $\delta$ being the drift quantity. From the whole Hamiltonian, the imperfect dynamic NOT and Hadamard gates can be constructed by taking the pulse area satisfying $\pi$ and $\frac {3\pi} {2}$, respectively, where during the process of constructing the Hadamard gate, the relative phase needs to be reduced by $\frac {\pi} {2}$ at the moment when the pulse area reaches $\pi$. In Fig. \ref{Fig3}, we present our numerical simulation result for both gates, which shows that the geometric gates are indeed less sensitive to qubit-frequency drift error.

\subsection{Composite scheme}

\begin{figure}[tbp]
\centering
\includegraphics[width=0.9\linewidth]{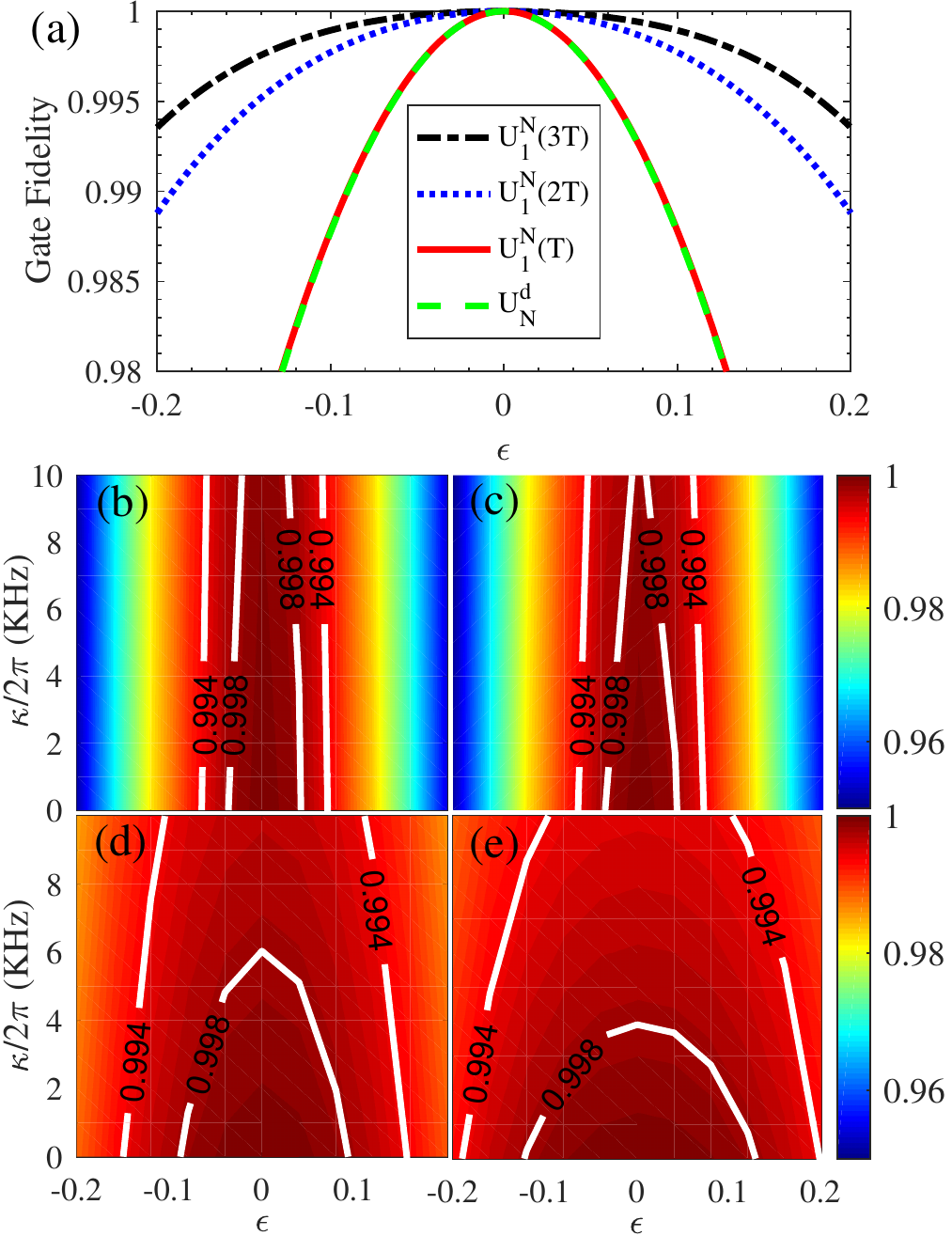}
\caption{ Gate fidelities under systematic errors. (a) Gate fidelities as a function of the error. Gate fidelities under the error and decoherence for dynamic $U_N^d$ and  geometric $U^N_1(\mathrm{T})$ gates are shown in (b) and (c), which have similar performance. Figures (d) and (e) show the results of the geometric composite gates $U^N_1(2\mathrm{T})$ and  $U^N_1(3\mathrm{T})$, which indicate the improved gate performance.}
  \label{Fig4}
\end{figure}

Next, we use the demonstrated composite-pulse scheme to improve the ability of our geometric gates in suppressing systematic error, which can further enhance the robustness of our geometric gates (the analytical calculation is present in Appendix B). Here, the systematic error we consider is in the form of $(1+\epsilon) \Omega(t)$ with $\epsilon$ being the error fraction, i.e., the deviation of driving strength, which causes the destruction of the exact $\pi$ pulse condition so that the cyclic evolution is no longer satisfied. We take the single-qubit gate $U_1(\mathrm{T})$ in Eq. (\ref{Eq5}) as the elementary gate and sequentially apply the elementary gate $\mathrm{n}$ times with $n>1$. Then, the following single-qubit geometric composite gate,
\begin{eqnarray}
\label{Eq8}
U_1(\mathrm{n} \mathrm{T})&\equiv& U^{\mathrm{n}}_1(\mathrm{T})
=e^{i\mathrm{n} \gamma}|\psi_+\rangle\langle \psi_+|+e^{-i\mathrm{n} \gamma}|\psi_-\rangle\langle \psi_-|,
\end{eqnarray}
can be achieved.

In order to reflect the advantage of the composite-pulses scheme, we first consider only the influence of the systematic error. Here, taking the NOT gate as an example and setting the parameters of the qubit as $\alpha=2\pi\times220$ MHz, $\Omega_0=2\pi\times40$ MHz. We select the geometric gate $U^N_1(\mathrm{T})$, the geometric composite gates $U^N_1(2\mathrm{T})$, $U^N_1(3\mathrm{T})$ and the dynamic gate $U_N^d$, which can be obtained by simple Rabi oscillation, as our comparison objects. In Fig. \ref{Fig4}(a), we plot the gate fidelities as a function of the systematic error, one can clearly find that the geometric composite gates $U^N_1(2\mathrm{T})$ and $U^N_1(3\mathrm{T})$ have better performance than the geometric gate $U^N_1(\mathrm{T})$ and the dynamic gate $U_N^d$ under the systematic error. In addition, we note that the geometric composite gates will inevitably require a longer operation time, and thus we need to comprehensively analyze the influence of both the decoherence and the systematic error, after neglecting the higher-order oscillating terms (corrected by DRAG). As shown in Fig. \ref{Fig4}(b) and (c), under decoherence, both the geometric and dynamical gates share similar performance in terms of the systematic error. However, we find that the geometric composite gates can surpass the dynamical gate, as shown in Fig. \ref{Fig4} (d) and (e). Meanwhile, considering the competition of the systematic and decoherence errors, $U^N_1(2\mathrm{T})$ is the best choice in our case.

\section{Nontrivial two-qubit geometric gates}

In this section, we consider the construction of the nontrivial two-qubit geometric gates. Therefore, combining the implemented arbitrary single-qubit gates, nonadiabatic GQC can be realized on the 2D square superconducting qubit lattice, where all the adjacent transmon qubits are capacitively coupled. Our two-qubit gates can be implemented in any two adjacent transmon qubits. In addition, similar to the single-qubit-gate case, in order to consider the effect of the intrinsic weak anharmonicity of the transmon qubits, we need to take the third energy level that is beyond the qubit states into account, as main leakages out of the qubit basis come from this level during the construction of two-qubit gates.

\begin{figure*}[tbp]
  \centering
  \includegraphics[width=0.85\linewidth]{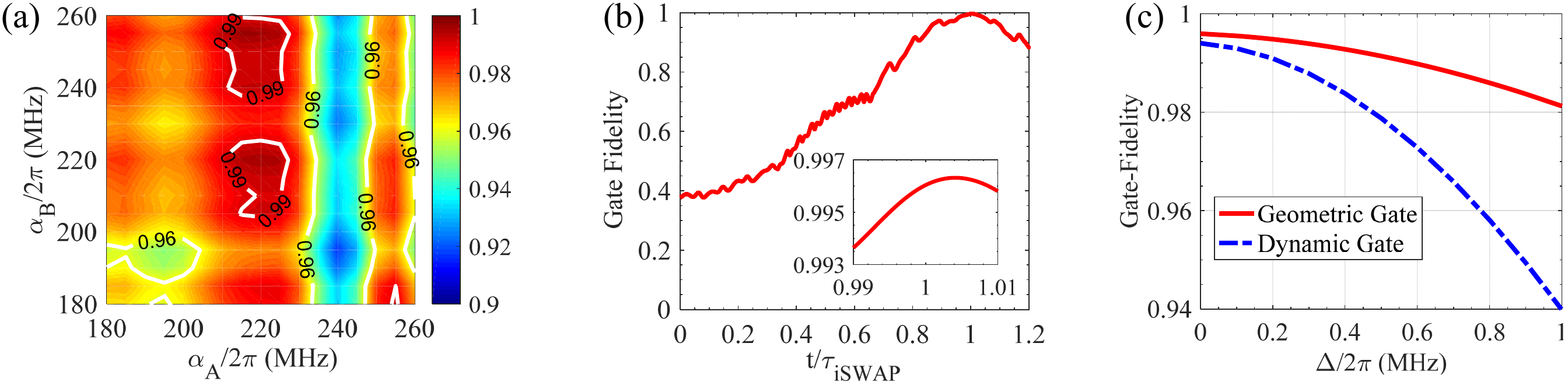}
  \caption{ (a) Gate fidelities as functions of the anharmonicities of the transmon qubits $\mathrm{T}_\mathrm{A}$ and $\mathrm{T}_\mathrm{B}$. (b) Dynamics of the gate fidelity of  the iSWAP geometric gate. The simulations are performed based on the interaction Hamiltonian in Eq. (\ref{Eq9}), i.e., without introducing any approximation. (c) Gate fidelities for both geometric and dynamic iSWAP two-qubit gates as a function of the qubit-frequency drift difference $\Delta$.}
  \label{Fig5}
\end{figure*}

\subsection{Parametrically tunable coupling}

The nontrivial two-qubit gates can be implemented on two capacitively coupled transmon qubits, e.g., the two denoting as $\mathrm{T}_\mathrm{A}$ and $\mathrm{T}_\mathrm{B}$, the details are shown in Fig. \ref{Fig1}(d). The Hamiltonian can be written as
\begin{eqnarray}
\label{Eq9}
H_c&=&\sum_{k=\mathrm{A},\mathrm{B}} \sum_{j=1}^2 [j\omega_k-(j-1)\alpha_k]\chi^k_j                                \notag\\
   &+& g_{_{\mathrm{AB}}} \prod_{k=\mathrm{A},\mathrm{B}} \left(\sum_{j=1}^2 \lambda_j\sigma^k_j \right)+ \mathrm{H.c.},
\end{eqnarray}
where $g_{_{\mathrm{AB}}}$ is the qubit-qubit coupling strength, and the identifier \emph{k} is used to distinguish difference qubits.

Usually, the qubit-frequency difference of adjacent transmon qubits and their coupling strength are fixed and not adjustable. To obtain tunable coupling between them \cite{TC5,TC6,TC7,TC8}, we add an ac driving on the transmon qubit $\mathrm{T}_\mathrm{B}$, experimentally induced by biasing the qubit by an ac magnetic flux, which results in periodically modulating this qubit's transition frequency in the form of
\begin{eqnarray}
\label{Eq10}
\omega_{_{\mathrm{B}}}(t)=\omega_{_{\mathrm{B}}}+\varepsilon_1 \sin(\nu_1 t+\varphi).
\end{eqnarray}
Moving into the interaction picture, the transformed Hamiltonian reads
\begin{eqnarray}
\label{Eq11}
H_I&=& g_{_{\mathrm{AB}}} \left\{ |10\rangle\langle 01|e^{i\Delta_1 t}e^{i\beta_1 \cos(\nu_1 t+\varphi)}\right. \notag \\
&+&\left.\sqrt{2}|11\rangle\langle 02|e^{i(\Delta_1+\alpha_{_{\mathrm{B}}}) t}e^{i\beta_1 \cos(\nu_1 t+\varphi)}\right. \\
&+&\left.\sqrt{2}|20\rangle\langle 11|e^{i(\Delta_1-\alpha_{_{\mathrm{A}}}) t}e^{i\beta_1 \cos(\nu_1 t+\varphi)} +\mathrm{H.c.}\right\},   \notag
\end{eqnarray}
where $\Delta_1=\omega_{_{\mathrm{A}}}-\omega_{_{\mathrm{B}}}$, $\beta_1=\varepsilon_1/\nu_1$, and $|mn\rangle=|m\rangle_{_{\mathrm{A}}}\otimes|n\rangle_{_{\mathrm{B}}}$. We can see that the resonant interaction can be induced from the above Hamiltonian in both the single- or two-excitation subspaces by a different choice of the driving frequency $\nu_1$; the corresponding energy level diagram is shown in Fig. \ref{Fig1}(e).

In recent experiments \cite{TC6,TC7}, the above parametrically tunable coupling is used to implement the two-qubit iSWAP dynamical gate $U_S^d$ in the single-excitation subspace $\{|10\rangle,|01\rangle\}$ by meeting the condition of
\begin{eqnarray}
\label{Eq12}
\Delta_1=n_1\nu_1,
\end{eqnarray}
where $n_1=\pm1,\pm2,\ldots$, and two kinds of two-qubit control-phase gates in the two-excitation subspaces $\{|11\rangle,|02\rangle\}$ and $\{|20\rangle,|11\rangle\}$ by meeting the condition of
\begin{eqnarray}
\label{Eq13}
\Delta_1+\alpha_{_\mathrm{B}}=n_2\nu_1, \quad \Delta_1-\alpha_{_\mathrm{A}}=n_3\nu_1
\end{eqnarray}
respectively, where $n_2,n_3=\pm1,\pm2,\ldots$. The experimental results indicate that the gate infidelities are $6\%$ and $9\%$ for the two-qubit for iSWAP and Control-phase gates, respectively. These gate infidelities mainly come from the influence of the third level of the transmons, due to the limited qubit anharmonicities. In the next subsection and in the next section dealing with DFS encoding, we show that the improvement of gate fidelities for the iSWAP and control-phase gates can be achieved through the optimization of qubit parameters. In particular, we can use this parametrically tunable coupling to complete the construction of high-fidelity two-qubit iSWAP and control-phase gates in a geometric way.

\subsection{Two-qubit geometric gates}

We consider the case of the parametric driving compensating the energy-splitting difference between the transmon qubits $\mathrm{T}_\mathrm{A}$ and $\mathrm{T}_\mathrm{B}$, i.e., meeting the condition $\Delta_1=\nu_1$ as given in Eq. (\ref{Eq12}). Then, using the Jacobi-Anger identity
\begin{eqnarray}
\label{Eq0}
\exp[i\beta_1 \cos(\nu_1 t+\varphi)]
=\sum^{\infty}_{m=-\infty} i^m J_m(\beta_1) \exp[im(\nu_1 t+\varphi)]    \notag
\end{eqnarray}
with $J_m(\beta_1)$ being Bessel functions of the first kind, and applying the rotating-wave approximation by neglecting the higher-order oscillating terms, we get an effectively resonant interaction Hamiltonian in the single-excitation subspace as
\begin{eqnarray}
\label{Eq14}
H_2= g^{\prime}_{_{\mathrm{AB}}} (|10\rangle\langle 01|e^{-i(\varphi-\frac {\pi} {2})}+ \mathrm{H.c.}),
\end{eqnarray}
where $g^{\prime}_{_{\mathrm{AB}}}=J_1(\beta_1)g_{_{\mathrm{AB}}}$.

In the same way as that of the single-qubit-gate case, we divide the entire evolution time $\tau$ into three parts to achieve geometric evolution, at intermediate time $\tau_1$, $\tau_2$ with pulse area, and relative phase $\varphi$, satisfying
\begin{eqnarray}
\label{Eq15}
g^{\prime}_{_{\mathrm{AB}}}\tau_1&=&\frac {\vartheta} {2},                       \quad \varphi,    \quad \quad \quad \quad \quad  t\in[0,\tau_1]          \notag\\
g^{\prime}_{_{\mathrm{AB}}}(\tau_2-\tau_1)&=& \frac {\pi} {2},                   \quad \varphi+\xi+\pi,                    \quad  t\in[\tau_1,\tau_2]     \notag\\
g^{\prime}_{_{\mathrm{AB}}}(\tau-\tau_2)&=&\frac {\pi} {2}-\frac {\vartheta} {2},\quad \varphi.      \quad \quad \quad            t\in[\tau_2,\tau]
\end{eqnarray}
At the final time $\tau$, the evolution operator can be expressed as
\begin{eqnarray}
\label{Eq16}
U_2(\tau)&=&U_2(\tau,\tau_2) U_2(\tau_2,\tau_1) U_2(\tau_1,0) \\
&=&\left(
\begin{array}{cccc}
 1 & 0 & 0 & 0 \\
 0 & \cos{\xi}-i\sin{\xi} \cos{\vartheta} & i\sin{\xi} \sin{\vartheta}e^{i\varphi} & 0 \\
 0 & i\sin{\xi} \sin{\vartheta}e^{-i\varphi} & \cos{\xi}+i\sin{\xi} \cos{\vartheta} & 0 \\
 0 & 0 & 0 & 1 \\
\end{array}
\right)     \notag
\end{eqnarray}
in the two-qubit subspace $\{|00\rangle,|01\rangle,|10\rangle,|11\rangle\}$.
Therefore, the two-qubit geometric gates can be achieved. In the case of $\vartheta=\pi/2$, $\varphi=0$, and $\xi=\pi/2$, the induced two-qubit geometric gate is an iSWAP gate
\begin{eqnarray}
\label{Eq17}
U_2^S=
\left(
\begin{array}{cccc}
 1 & 0 & 0 & 0 \\
 0 & 0 & i & 0 \\
 0 & i & 0 & 0 \\
 0 & 0 & 0 & 1 \\
\end{array}
\right),
\end{eqnarray}
which is a nontrivial two-qubit entangling gate for universal GQC, since it can transform the product state $(|0\rangle_{_\mathrm{A}}+|1\rangle_{_\mathrm{A}})\otimes (|0\rangle_{_\mathrm{B}}+|1\rangle_{_\mathrm{B}})/2$ into an entangled state $(|00\rangle+i|01\rangle +i|10\rangle+|11\rangle)/2$ after the gate operation.

To fully evaluate the performance of the implemented two-qubit gate, for the general initial state $|\psi_2\rangle=(\cos\vartheta_1|0\rangle_{_\mathrm{A}}+\sin\vartheta_1|1\rangle_{_\mathrm{A}})\otimes
(\cos\vartheta_2|0\rangle_{_\mathrm{B}}+\sin\vartheta_2|1\rangle_{_\mathrm{B}})$, we define the two-qubit gate fidelity as $F^G_S=\frac {1} {4\pi^2}\int_0^{2\pi} \int_0^{2\pi} \langle \psi_{f_S}|\rho_2|\psi_{f_S}\rangle d\vartheta_1d\vartheta_2$ with $|\psi_{f_S}\rangle=U_2^S|\psi_2\rangle$ being the ideal final state. Set $\Delta_1=2\pi\times146$ MHz, $g_{_{\mathrm{AB}}}=2\pi\times8$ MHz and $\beta_1=\varepsilon_1/\nu_1=\varepsilon_1/\Delta_1\approx2.1$. In Fig. \ref{Fig5}(a), we plot the gate fidelities as functions of the anharmonicities of the transmon qubits $\mathrm{T}_\mathrm{A}$ and $\mathrm{T}_\mathrm{B}$ with the uniform decoherence rate being $\kappa=2\pi\times4$ KHz. When $\alpha_{_\mathrm{A}}=2\pi\times220$ MHz and $\alpha_{_\mathrm{B}}=2\pi\times255$ MHz, the gate fidelity can be as high as $99.61\%$, as shown in Fig. \ref{Fig5}(b). Meanwhile, we find that the gate infidelities are in the range of $3-5\%$ with parameters in the experiments of Refs. \cite{TC6,TC7}, which is in good agreement with their experimental results when taking the different decoherence rates into consideration.

In addition, to further show the robust feature of the geometric iSWAP gate, we take the error caused by the qubit-frequency drift as an example to analyze its influence for the geometric and dynamic iSWAP gate performance. When the qubit-frequency drifts of the two qubits do not synchronous, this error turns $\Delta_1$ of the Hamiltonian in Eq. (\ref{Eq11}) into $(\Delta_1+\Delta)$ with $\Delta$ being the drift difference. The simulation result in Fig. \ref{Fig5}(c) does show that the geometric gate is less sensitive to this drift error.

\subsection{Composite scheme}

\begin{figure}[tbp]
  \centering
  \includegraphics[width=0.9\linewidth]{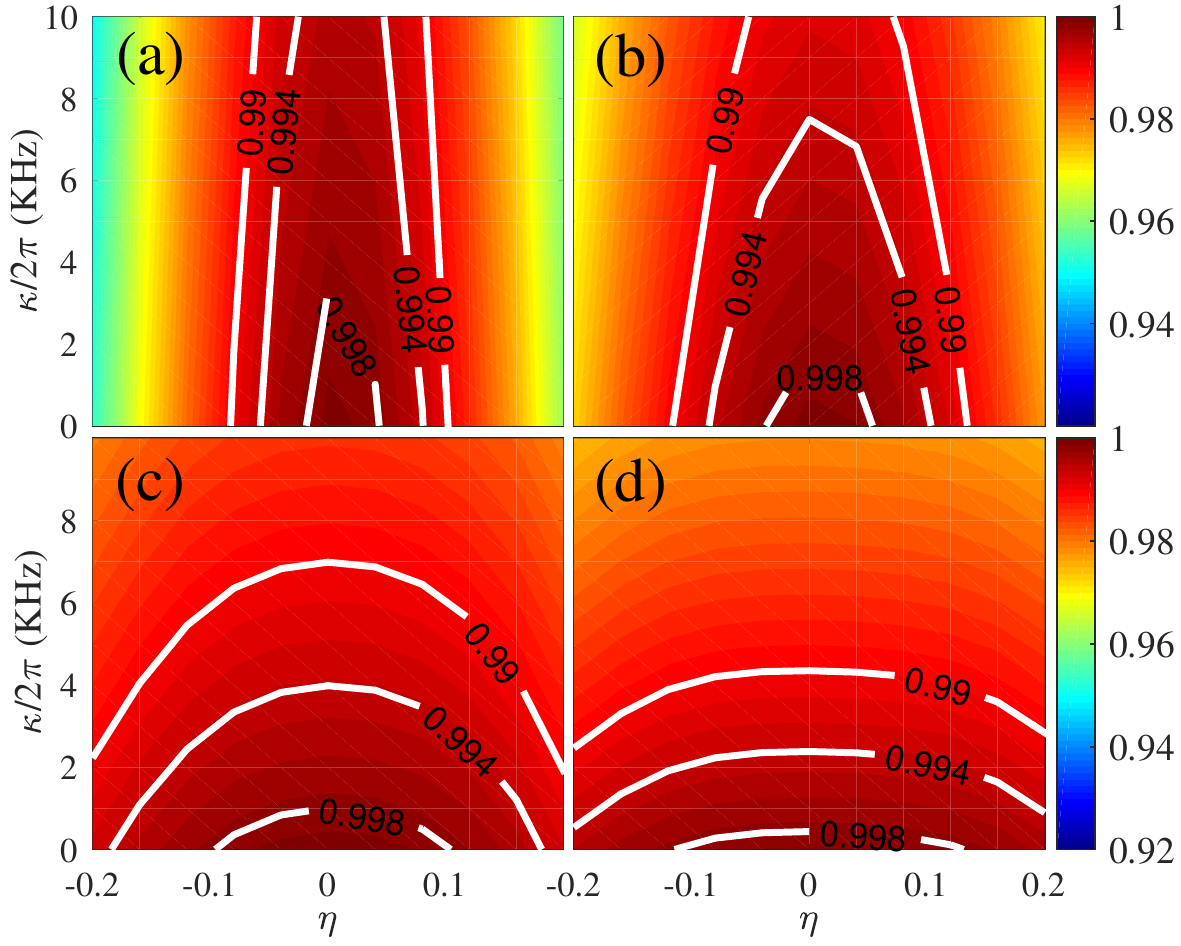}
  \caption{Gate fidelities versus the decoherence and systematic error. The results of the dynamic gate $U_S^d$ and the geometric gate $U_2^S(\tau)$ are shown in (a) and (b), respectively. Figures (c) and (d) show the geometric composite gates $U_2^S(2\tau)$, $U_2^S(3\tau)$, respectively.}
  \label{Fig6}
\end{figure}

The composite-pulse scheme can also be applied to our two-qubit geometric gates, where the systematic error can be suppressed. Here we define the form of systematic error as $(1+\eta)g_{_{\mathrm{AB}}}$ with $\eta$ being the error fraction, which denotes the deviation of the qubit-qubit coupling strength. We take the two-qubit gate $U_2(\tau)$ in Eq. (\ref{Eq16}) as the elementary gate, sequentially apply the elementary gate m times and achieve difference kinds of two-qubit geometric composite gate $U_2(\mathrm{m}\tau)$. In the following, we take the iSWAP two-qubit gate as an example and select the geometric gate $U_2^S(\tau)$, the geometric composite gates $U_2^S(2\tau)$, $U_2^S(3\tau)$ and the dynamic gate $U_S^d$ as our comparison objects under the above optimal parameters. As shown in Fig. \ref{Fig6}, taking into account both the decoherence and the systematic error, based on the original Hamiltonian in Eq. (\ref{Eq9}), we can find the superiority of the geometric composite gates in suppressing the systematic error, and the best performance is the $U_2^S(2\tau)$ gate.

\section{GQC with DFS encoding}

In order to suppress collective dephasing noise \cite{DFS1,DFS2,DFS3} and further improve the precision of the quantum-gate control, we now propose to realize nonadiabatic GQC in DFS on the 2D square qubit lattice, as shown in Fig. \ref{Fig1}(a), where we only use two physical qubits to encode a DFS logical qubit, with minimal qubit resource requirement. To avoid confusion with physical qubits, we use the subscript L to denote logical qubits in the following. The definition of the logical qubit states of the 2D DFS is
\begin{eqnarray}
\label{Eq18}
|0\rangle_L=|10\rangle,\ \  |1\rangle_L=|01\rangle,
\end{eqnarray}
where $|mn\rangle=|m\rangle_{_\mathrm{A}}\otimes|n\rangle_{_\mathrm{C}}$.

\subsection{Single-logical qubit geometric gates}

Firstly, we turn to the implementation of universal single-logical qubit geometric gates by parametrically tunable coupling between two capacitively coupled transmon qubits $\mathrm{T}_\mathrm{A}$ and $\mathrm{T}_\mathrm{C}$, which can be achieved by adding a parametric driving on the transmon qubit $\mathrm{T}_\mathrm{C}$ so that its frequency oscillation as $\omega_{_{\mathrm{C}}}(t)=\omega_{_{\mathrm{C}}}+\varepsilon_2 \sin(\nu_2 t+\phi_L)$. Here we can take them as a unit to encode our logic qubit. The effective coupling Hamiltonian, which is similar to Eq. (\ref{Eq14}), reads
\begin{eqnarray}
\label{Eq19}
H_{L_1}= g^{\prime}_{_{\mathrm{AC}}} [|0\rangle_L\langle 1|e^{-i(\phi_{_L}-\frac {\pi} {2})}+ \mathrm{H.c.}],
\end{eqnarray}
where $g^{\prime}_{_{\mathrm{AC}}}=J_1(\beta_2)g_{_{\mathrm{AC}}}$ with $\beta_2=\varepsilon_2/\nu_2$. As our DFS encoding is within the single-excitation subspace, the encoded quantum information does not leak to other subspaces.

To achieve geometric evolution, we divide the entire evolution time $\mathrm{T}'$ into three parts, at intermediate time $\mathrm{T}'_1$ and $\mathrm{T}'_2$, with pulse area and relative phase $\phi_{_L}$ satisfying
\begin{eqnarray}
\label{Eq20}
g^{\prime}_{_{\mathrm{AC}}}\mathrm{T}'_1&=&\frac {\theta_L} {2},                              \quad \phi_{_L},
\quad \quad \quad \quad \quad t\in[0,\mathrm{T}'_1], \notag\\
g^{\prime}_{_{\mathrm{AC}}}(\mathrm{T}'_2-\mathrm{T}'_1)&=& \frac {\pi} {2},                  \quad \phi_{_L}+\gamma_{_L}+\pi,
\quad t\in[\mathrm{T}'_1,\mathrm{T}'_2],\notag\\
g^{\prime}_{_{\mathrm{AC}}}(\mathrm{T}'-\mathrm{T}'_2)&=&\frac {\pi} {2}-\frac {\theta_L} {2},\quad \phi_{_L},
\quad \quad \quad t\in[\mathrm{T}'_2,\mathrm{T}'].\quad
\end{eqnarray}
At the final time $\mathrm{T}'$, the evolution operator reads
\begin{eqnarray}
\label{Eq21}
U_{L_1}(\mathrm{T}')&=&U_{L_1}(\mathrm{T}',\mathrm{T}'_2) U_{L_1}(\mathrm{T}'_2,\mathrm{T}'_1) U_{L_1}(\mathrm{T}'_1,0) \\
&=&\cos{\gamma_{_L}}+i\sin{\gamma_{_L}}\left(
\begin{array}{cccc}
 \cos{\theta_L} & \sin{\theta_L}e^{-i\phi_{_L}} \\
 \sin{\theta_L}e^{i\phi_{_L}} & -\cos{\theta_L}
\end{array} \right), \notag
\end{eqnarray}
in the two-dimensional DFS $S_1=\{|0\rangle_L,|1\rangle_L\}$, from which universal nonadiabatic single-qubit geometric gates in DFS can be achieved. The geometric nature of this gate can be proved similar to that of the single-qubit case in Eq. (\ref{Eq3}).

To further analyze the performance of the single-logical qubit gates, here we choose the NOT and Hadamard gates as two typical examples, which correspond to the same $\phi_{_L}=0$ and $\gamma_{_L}=\pi/2$, with different $\theta_{_L}$, $\theta_{_L}^N=\pi/2$, and $\theta_{_L}^H=\pi/4$ for the NOT and Hadamard gates, respectively. Here, the anharmonicities of the transmon qubits $\mathrm{T}_\mathrm{A}$ and $\mathrm{T}_\mathrm{C}$ are $\alpha_{_\mathrm{A}}=2\pi\times220$ MHz and $\alpha_{_\mathrm{C}}=2\pi\times245$ MHz. Meanwhile, we set the parametric driving frequency equal to the frequency difference of the two transmon qubits $\mathrm{T}_\mathrm{A}$ and $\mathrm{T}_\mathrm{C}$, i.e., $\nu_2=\Delta_2=\omega_{_{\mathrm{A}}}-\omega_{_{\mathrm{C}}}=2\pi\times165$ MHz, so that the effectively resonant interaction is induced in the single-excitation subspace. Set the decoherence rates of all the physical qubits as $\kappa=2\pi\times4$ KHz, for $\beta_2=\varepsilon_2/\nu_2 \approx2.2$ and the qubit-qubit coupling strength $g_{_{\mathrm{AC}}}=2\pi\times20$ MHz, the gate fidelities of the NOT and Hadamard gates can be as high as $F_N^L=99.81\%$ and $F_H^L=99.87\%$, as shown in Fig. \ref{Fig7}(a).

\subsection{Two-logical qubit geometric gates}

We next proceed to the construction of the two-logical qubit control-phase geometric gates. As shown in Fig. \ref{Fig1}(a), any two capacitively coupled qubits on the 2D square superconducting qubit lattice can be treated as a unit element, which is used to encode a logical qubit for our DFS encoding. Here, we choose the transmon qubits $\mathrm{T}_\mathrm{A}$, $\mathrm{T}_\mathrm{C}$ and  $\mathrm{T}_\mathrm{B}$, $\mathrm{T}_\mathrm{D}$ to encode the first and second logical qubits, respectively. In this case, a four-dimensional DFS exists,
\begin{eqnarray}
\label{Eq22}
S_2=\{& &|00\rangle_L=|1010\rangle,\   \ |01\rangle_L=|1001\rangle,   \notag\\
      & &|10\rangle_L=|0110\rangle,\   \ |11\rangle_L=|0101\rangle\},
\end{eqnarray}
where $|mnm'n'\rangle=|m\rangle_\mathrm{A}\otimes|n\rangle_\mathrm{C}\otimes|m'\rangle_\mathrm{B} \otimes|n'\rangle_\mathrm{D}$.

\begin{figure}[tbp]
  \centering
  \includegraphics[width=\linewidth]{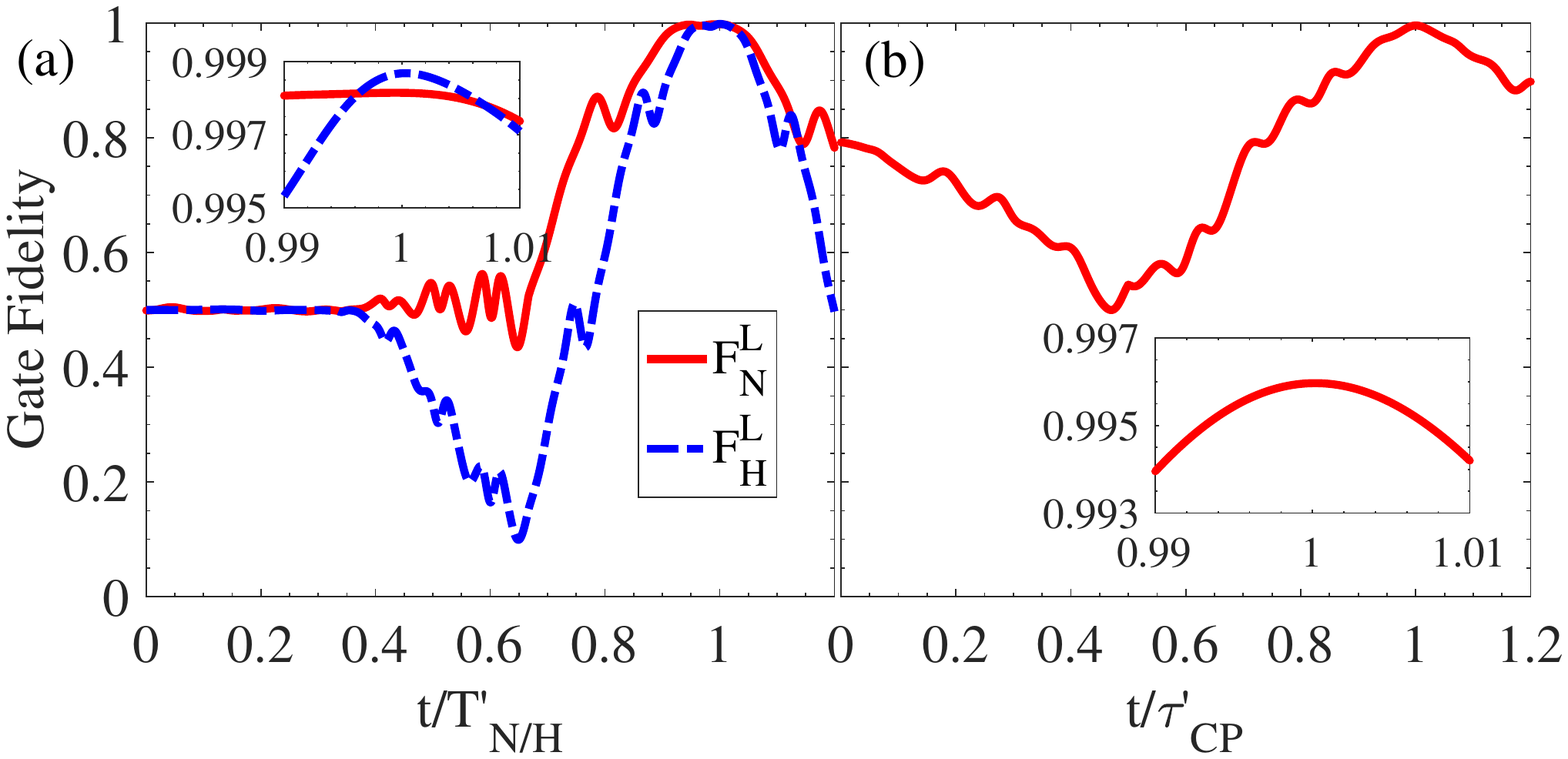}
  \caption{ Dynamics of the gate fidelities of (a) single-logical qubit NOT and Hadamard gates, and (b) two-logical qubit control-phase gate. Numerical simulations are based on the original interaction Hamiltonian without any approximation, which thus also verifies our analytical results.}
  \label{Fig7}
\end{figure}

For the construction of two-logical qubit geometric gates, we only need the parametrically tunable coupling between two physical qubits, i.e., one from each logical qubit, namely $\mathrm{T}_\mathrm{C}$ and $\mathrm{T}_\mathrm{D}$, by adding an ac driving on the qubit $\mathrm{T}_\mathrm{C}$ so that its frequency oscillation $\omega'_{_{\mathrm{C}}}(t)=\omega_{_{\mathrm{C}}}+\varepsilon_3 \sin(\nu_3 t+\varphi_{_L})$ can be obtained. The above two-logical qubits are arranged in the horizontal direction in the 2D lattice of Fig. \ref{Fig1}(a), for the case of two-logical qubits arranged in the vertical direction in the 2D lattice, we can use the interaction between the nearest two qubits to induce our wanted geometric gate, namely $\mathrm{T}_\mathrm{C}$ and $\mathrm{T}_\mathrm{E}$. The interaction Hamiltonian can be written as
\begin{eqnarray}
\label{Eq23}
H'_I&=& g_{_{\mathrm{CD}}} \left\{ |01\rangle_{_{\mathrm{CD}}}\langle 10|e^{i\Delta_3 t}e^{i\beta_3 \cos(\nu_3 t+\varphi_{_L})}\right.    \notag \\
&+&\left.\sqrt{2}|02\rangle_{_{\mathrm{CD}}}\langle 11|e^{i(\Delta_3-\alpha_{_{\mathrm{D}}}) t}e^{i\beta_3 \cos(\nu_3 t+\varphi_{_L})}\right. \\
&+&\left.\sqrt{2}|11\rangle_{_{\mathrm{CD}}}\langle 20|e^{i(\Delta_3+\alpha_{_{\mathrm{C}}}) t}e^{i\beta_3 \cos(\nu_3 t+\varphi_{_L})} +\mathrm{H.c.}\right\},  \notag
\end{eqnarray}
where $\Delta_3=\omega_{_{\mathrm{D}}}-\omega_{_{\mathrm{C}}}$, $\beta_3=\varepsilon_3/\nu_3$. The resonant situation in the two-excitation subspace of $\{|11\rangle_{_{\mathrm{CD}}},|02\rangle_{_{\mathrm{CD}}}\}$ can be achieved by setting $\nu_3=\alpha_{_{\mathrm{D}}}-\Delta_3$. Then, we apply the rotating-wave approximation by neglecting the higher-order oscillating terms, we can obtain an effectively resonant interacting Hamiltonian  as
\begin{eqnarray}
\label{Eq24}
H_{L_2}= g^{\prime}_{_{\mathrm{CD}}} (|11\rangle_L\langle a|e^{-i(\varphi_{_L}+ \frac{\pi} {2})}+ \mathrm{H.c.}),
\end{eqnarray}
where $g^{\prime}_{_{\mathrm{CD}}}=\sqrt{2}J_1(\beta_3)g_{_{\mathrm{CD}}}$,  and $|a\rangle_L=|0002\rangle$, which is regarded as an auxiliary state here.

In order to achieve the construction of the two-logical qubit control-phase gates, we divide the total evolution time $\tau'$ into two equal parts. For the first stage $t\in[0,\tau'/2]$, the relative phase $\varphi_{_L}=\pi$ in the Hamiltonian $H_{L_2}$. At the moment $\tau'/2$, we change the relative phase to $\varphi_{_L}=\xi_{_L}$. For the second stage $t\in[\tau'/2,\tau']$, the relative phase remains at $\varphi_{_L}=\xi_{_L}$. The whole evolution of logical states forms an orange-slice-shaped evolution path. After undergoing this cyclic evolution path, the logical state $|11\rangle_L$ can obtain a purely geometric phase $-\xi_{_L}$ as the parallel-transport condition is satisfied, i.e.,
\begin{eqnarray}
\label{Eq25}
_L\langle11|U_{L_2}^\dag(t)H_{L_2}U_{L_2}(t)|11\rangle_L=0.
\end{eqnarray}
The evolution operator $U_{L_2}(t)$ at the final time $\tau'$ is
\begin{eqnarray}
\label{Eq26}
U_{L_2}(\tau')&=&U_2(\tau',\tau'_1) U_2(\tau'_1,0)       \notag\\
&=&\left(
\begin{array}{cccc}
 1 & 0 & 0 & 0 \\
 0 & 1 & 0 & 0 \\
 0 & 0 & 1 & 0 \\
 0 & 0 & 0 & e^{-i\xi_{_L}} \\
\end{array}
\right)
\end{eqnarray}
in the DFS $S_2$. Thus, the two-logical qubit control-phase geometric gates can be achieved. To further evaluate the performance of these gates, we take $U_{\mathrm{CP}}=\mathrm{diag}\{1,1,1, e^{-i\pi/2} \}$ as a typical example by taking $\xi_{_L}=\pi/2$. Here we set the parameters of the qubit $\mathrm{T}_\mathrm{D}$ as $\alpha_{_\mathrm{D}}=2\pi\times 200$ MHz, $\kappa=2\pi\times4$ KHz and the coupling strength is $g_{_{\mathrm{CD}}}=2\pi\times10$ MHz. For $\Delta_3=2\pi\times130$ MHz, $\nu_3=\alpha_{_\mathrm{D}}-\Delta_3=2\pi\times70$ MHz and $\beta_3\simeq2.1$, the gate fidelity of $U_{\mathrm{CP}}$ can reach $99.60\%$, as shown in Fig. \ref{Fig7}(b). It is worth emphasizing that the composite-pulse scheme in the previous section can also be applied to this DFS-encoding scheme of GQC.

\section{Conclusion}

In summary, we propose to implement nonadiabatic GQC on the 2D square superconducting qubit lattice consisting of capacitively coupled transmon qubits, by using parametrically tunable coupling between two qubits. During the geometric gate operations, we achieve independent manipulation of the qubit states, without introducing any auxiliary state. Moreover, our scheme can be achieved by effectively tunable all-resonant interaction, thus leading to high-fidelity and more robust geometric quantum-composite gates. In addition, nonadiabatic GQC in DFS, based on parametrically tunable resonant coupling between two transmon qubits, can also be implemented, without consulting multiqubit interactions. Moreover, in our scheme, we only use two physical qubits to construct the DFS, which possesses minimal qubit resource requirement. Therefore, our scheme provides a promising way to achieve high-fidelity GQC.

\begin{acknowledgements}
We thank B.-J. Liu, P. Z. Zhao, and Dr J. Liu for suggestions and discussions. This work is supported in part by the National Natural Science Foundation of China (Grant No. 11874156), the National Key R\&D Program of China (Grant No. 2016YFA0301803), and the ``Challenge Cup'' Golden Seed Training Project from SCNU (Grant No. 18WDKB01).
\end{acknowledgements}

\appendix
\section{The transmon qubit and its efficient control}

In the main text, we use the two lowest levels of the transmon as our qubit states, as depicted in Fig. \ref{Fig1}(c). These levels are separated in energy by $\hbar\omega$ with $\omega$ being the transition frequency. Due to the weak anharmonicity of the transmon qubit, when we intend to induce a resonant driving on the two lowest levels with a microwave field, it can also stimulate the transitions among the higher excited states in a dispersive way, as all the sequential transitions are allowed. However, the leakage out of our computational basis are mainly attributed to the third level. To this end, the Hamiltonian for a transmon is
\begin{eqnarray}
\label{EqA1}
H(t)&=&\sum_{j=1,2} [j\omega-(j-1)\alpha]\chi_j\notag\\
&+& \frac{1}{2}\sum_{j=1,2}\left[\Omega(t)\lambda_j\sigma_j e^{i\omega_d t-i\phi}+ \mathrm{H.c.} \right],
\end{eqnarray}
where $\chi_{j'}=|j'\rangle\langle j'|$ is the projector for the $j'$th level with $j'\in\{0,1,2\}$ denoting the three relevant levels,  $\sigma_j=|j-1\rangle\langle j|$ with $j\in\{1,2\}$ is the standard lower operator and the associated transition frequency is $[j\omega-(j-1)\alpha]$ with $\alpha$ being the intrinsic anharmonicity of the transmon qubit; $\Omega(t)$, $\omega_d$, and $\phi$ are the driving strength, frequency, and phase of the microwave field, respectively. The constant $\lambda_j$ weighs the relative strength of the $|j\rangle \leftrightarrow |j-1\rangle$ transition determined by the dipolar transition elements, with $\lambda_1=1$ and $\lambda_2=\sqrt{2}$.

To get a clear analysis for the interaction, we move to a rotating frame with respect to the driving frequency $\omega_d$. When $\omega_d=\omega$, the transformed Hamiltonian reads
\begin{eqnarray}
\label{EqA2}
H_t(t)=-\alpha\chi_2+\frac {1} {2} \left[\Omega(t)(\sigma_1+\sqrt{2}\sigma_2)e^{-i\phi}+\mathrm{H.c.} \right].\quad
\end{eqnarray}
That is, when the microwave field is applied to the two lowest levels of the qubit, it will also induce interaction between the second and third levels, resulting in the leakage error. We want to emphasize that this leakage error may be absent in natural atoms due to the atomic selection rule.

\begin{figure}[tbp]
  \centering
  \includegraphics[width=0.9\linewidth]{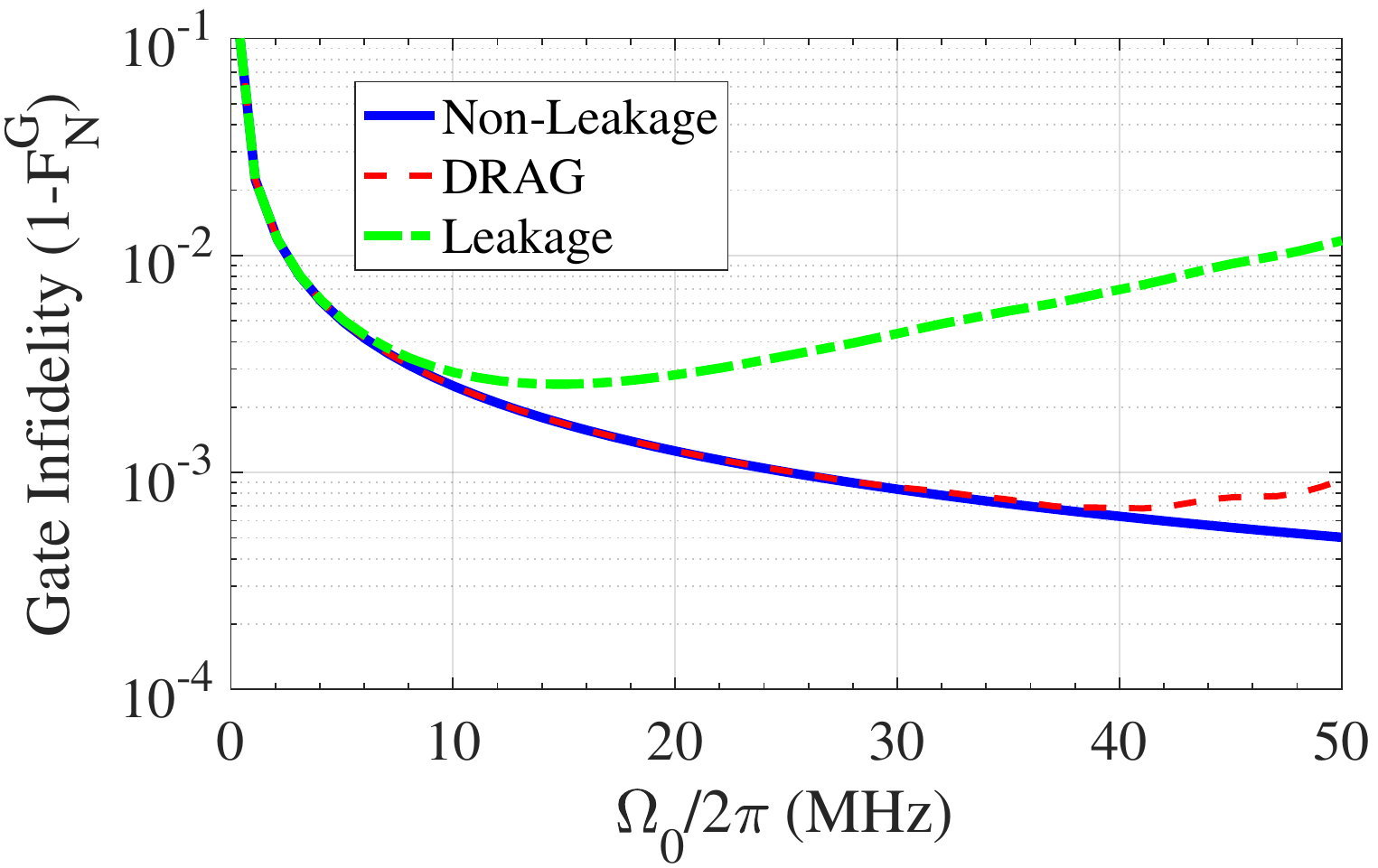}
  \caption{Different gate infidelities as a function of $\Omega_0$.}
  \label{Fig8}
\end{figure}

To quantitatively show how this leakage influences an intended gate operation, here we choose the geometric NOT gate as an typical example, which can be induced from Eq. (\ref{EqA1}) by setting the pulse area and relative phase $\phi$ satisfying Eq. (\ref{Eq2}). Meanwhile, we use a simple pulse shape for $\Omega(t)$ as
\begin{eqnarray}
\label{EqA3}
\Omega(t)=\Omega_0 \sin^2\left(\pi t/\mathrm{T}\right),
\end{eqnarray}
for $t\in [0, \mathrm{T}]$ with $\mathrm{T}$ being the gate duration. In addition, for practical physical implementation, the decoherence process is unavoidable. Therefore, we also include the decoherence effect by numerical simulation of the master equation of
\begin{eqnarray}
\label{EqA4}
\dot\rho_1&=& -i[H_t(t), \rho_1]+\sum_{j=1}^2\left[ \frac {\kappa^j_-} {2}\mathscr{L}(\sigma_j)+ \frac {\kappa^j_z} {2}\mathscr{L}(\chi_j) \right], \notag
\end{eqnarray}
where $\rho_1$ is the density matrix of the transmon, $\mathscr{L}(\mathcal{A})=2\mathcal{A}\rho_1
\mathcal{A}^\dagger-\mathcal{A}^\dagger \mathcal{A} \rho_1 -\rho_1 \mathcal{A}^\dagger \mathcal{A}$ is the Lindblad operator for operator $\mathcal{A}$, and $\kappa^j_-$, $\kappa^j_z$ are the relaxation and dephasing rates of the transmon, respectively. In our simulation, we choose the parameters from the state-of-the-art experiments in Refs. \cite{DRAGexperiment1,DRAGexperiment2}, that is $\alpha=2\pi\times 220$ MHz and $\kappa^1_-=\kappa^1_z=\frac {1} {2} \kappa^2_-=\frac {1} {2} \kappa^2_z=\kappa=2\pi\times 4$ KHz. In Fig. \ref{Fig8}, we plot the gate infidelity ($1-F_N^G$) as a function of the driving amplitude $\Omega_0$  in the absence (blue solid line) and presence (green dash-dot line) of the leakage error. During the gate operation, there are two competing factors for the gate fidelity, i.e., the decoherence and the leakage. For larger driving amplitude $\Omega_0$, the gate speed will be faster and the decoherence will induce less gate infidelity, while the dispersive coupling to the third level will lead to more gate infidelity. Finally, considering both  decoherence and leakage, we find that the maximum gate fidelity  is about $99.74\%$.

For large scale fault-tolerant quantum computation, high-fidelity quantum gates are preferred. Thus we want to outpace the above maximum fidelity in the presence of the leakage error, i.e., achieving independent manipulation of the  qubit states and removing the influence of the third level. For this purpose, we use the recent theoretical exploration of derivative removal via adiabatic gate (DRAG) \cite{DRAG1,DRAG2} to suppress the leakage error by correcting pulse shape in Eq. (\ref{EqA3}) as
\begin{eqnarray}
\label{EqA5}
\Omega_D(t)=\Omega(t)+i\frac {\dot{\Omega}(t)} {2\alpha}.
\end{eqnarray}
As shown in Fig. \ref{Fig8}, the gate infidelity can be suppressed to approach the case without leakage. Therefore, the DRAG correction can effectively suppress the leakage error, and the optimized gate fidelity is 99.93\% when $\Omega_0/2\pi=40$ MHz.

\section{Composite scheme}

Here, we present the analytical calculation to show the influence of systematic error on the gate fidelity and the advantage of the composite-pulse scheme in suppressing systematic error. In the presence of the systematic error, the Hamiltonian $H_1(t)$ for single-qubit gates in Eq. (\ref{Eq1}) turns to
\begin{eqnarray}
\label{EqB1}
H'_1(t)=
\frac {1} {2}\Omega'(t) \left(
\begin{array}{cccc}
 0 & e^{-i\phi} \\
 e^{i\phi} & 0
\end{array}
\right),
\end{eqnarray}
where $\Omega'(t)=(1+\epsilon)\Omega(t)=\mu\Omega(t)$ with $\epsilon$ being the error fraction, satisfying $|\epsilon|\ll1$.

After the geometric operation, the evolution operator under the influence of systematic error is calculated to be
\begin{eqnarray}
\label{EqB2}
U^\epsilon_1(\mathrm{T})&=&U^\epsilon_1(\mathrm{T},\mathrm{T}_2) U^\epsilon_1(\mathrm{T}_2,\mathrm{T}_1) U^\epsilon_1(\mathrm{T}_1,0) \notag\\
&=&\left[
\begin{array}{cccc}
 \cos\frac {\mu(\pi-\theta)} {2} & \sin\frac {\mu(\pi-\theta)} {2} e^{-i\phi} \\
 -\sin\frac {\mu(\pi-\theta)} {2} e^{i\phi} & \cos\frac {\mu(\pi-\theta)} {2}
\end{array}
\right] \notag\\
& &\times\left[
\begin{array}{cccc}
 \cos\frac {\mu\pi} {2}  & -\sin \frac {\mu\pi} {2} e^{-i(\phi+\gamma)} \\
 \sin \frac {\mu\pi} {2} e^{i(\phi+\gamma)} & \cos\frac {\mu\pi} {2}
\end{array}
\right]  \notag\\
& &\times\left(
\begin{array}{cccc}
 \cos\frac {\mu\theta} {2}  & \sin\frac {\mu\theta} {2} e^{-i\phi} \\
 -\sin\frac {\mu\theta} {2} e^{i\phi} & \cos \frac {\mu\theta} {2}
\end{array}
\right).
\end{eqnarray}

Next, we sequentially apply the decomposition and then the Taylor expansion formula of the trigonometric functions to Eq. (\ref{EqB2}), up to $\epsilon^2$, which are the main deviations here. According to the form of the elementary gate in the presence of systematic error, the corresponding composite gates are in the form of $U^\epsilon_1(2\mathrm{T})=U^\epsilon_1(\mathrm{T})U^\epsilon_1(\mathrm{T})$ and $U^\epsilon_1(3\mathrm{T})=U^\epsilon_1(\mathrm{T})U^\epsilon_1(\mathrm{T})U^\epsilon_1(\mathrm{T})$. To further compare the sensitivity of elementary gate and composite gates to systematic error, we calculate the fidelity of these gates by using the formula \cite{U}
\begin{eqnarray}
\label{EqB5}
F_{U^\epsilon}=\frac {U^{\dag}U^\epsilon} {\mathrm{Tr}(U^{\dag}U)},
\end{eqnarray}
where $U$ and $U^\epsilon$ represent the desired gate and the gate affected by the systematic error, respectively. In order to make the compared gate be the same, for the gate $U^\epsilon_1(\mathrm{T})$ and the composite gates $U^\epsilon_1(2\mathrm{T})$, $U^\epsilon_1(3\mathrm{T})$, we set the parameter $\gamma$ to be $\pi/2, \pi/4, \pi/6$, respectively. Through analytical calculations from Eq. (\ref{EqB5}), we can get
\begin{eqnarray}
\label{EqB6}
F_{U^\epsilon_1(\mathrm{T})}&\approx&1+\frac {1} {2}\epsilon^2\pi\theta-\frac {1} {2}\epsilon^2\theta^2-\frac {1} {4}\epsilon^2\pi^2, \notag\\
F_{U^\epsilon_1(2\mathrm{T})}&\approx&1+\frac {1} {2}\epsilon^2\pi\theta-\frac {1} {2}\epsilon^2\theta^2-(\frac {1} {2}-\frac {1} {2\sqrt{2}})\epsilon^2\pi^2, \notag\\
F_{U^\epsilon_1(3\mathrm{T})}&\approx&1+\frac {1} {2}\epsilon^2\pi\theta-\frac {1} {2}\epsilon^2\theta^2-(1-\frac {\sqrt{3}} {2})\epsilon^2\pi^2.
\end{eqnarray}
Comparing the above analytical results, we obtain
\begin{eqnarray}
\label{EqB7a8}
&& F_{U^\epsilon_1(\mathrm{2T})}-F_{U^\epsilon_1(\mathrm{T})}\approx\epsilon^2,\\
&& F_{U^\epsilon_1(\mathrm{3T})}-F_{U^\epsilon_1(\mathrm{T})}\approx1.2\epsilon^2.
\end{eqnarray}
Therefore, the composite-pulse scheme can strengthen the robustness of geometric gate against the systematic error. A similar discussion is also valid for the case of two-qubit geometric composite gates.

\end{document}